\documentclass[journal, twocolumn]{IEEEtran}
\IEEEoverridecommandlockouts
\usepackage[ruled,linesnumbered,vlined]{algorithm2e}
\usepackage{multirow}
\usepackage{makecell}
\usepackage{caption}
\usepackage{graphicx}
\usepackage{float}
\usepackage{pdfpages}
\usepackage{tabularx}
\usepackage{amsmath}
\usepackage{amsfonts,amssymb}
\usepackage{stmaryrd}
\usepackage{diagbox}
\usepackage[ruled]{algorithm2e}

\newtheorem{definition}{\textbf{Definition}}
\newtheorem{example}{\textbf{Example}}
\newtheorem{proposition}{\textbf{Proposition}}

\usepackage{cite}
\usepackage{bm}
\usepackage{xcolor}

\usepackage{CJK}
\usepackage{indentfirst}
\usepackage{cases}
\usepackage{subeqnarray}
\usepackage{lettrine}
\SetKwFor{ForAll}{for all}{do}{end for}

\usepackage{booktabs}
\usepackage{siunitx}
\usepackage{graphicx}
\usepackage{subcaption}

\begin{document}

\title{Dynamic Scheduling for Flexible Manufacturing Systems Based on Multi-Agent Deep Reinforcement Learning and Petri Nets}

%\author{Zhou He,~\IEEEmembership{Senior Member,~IEEE}, Ning Li, Ruotian Liu, Liang Li and Carla Seatzu,~\IEEEmembership{Senior Member,~IEEE}
\author{Zhou He, Ning Li, Ruotian Liu, Liang Li and Carla Seatzu,\thanks{ This work was supported in part by the National Natural Science Foundation of China under Grant nos. 62373234, 62373132, and 62303359, in part by Shaanxi Provincial Natural Science Foundation under Grant 2025ZC-KJXX-42, in part by the Natural Science Foundation of Hebei Province under Grant F2025201023, in part by the Natural Science Foundation of Hubei Province under Grant 2026AFB500, in part by the S\&T Program of Shijiazhuang under Grant 241791367A, in part by the Excellent Youth Research Innovation Team of Hebei University under Grant QNTD202411, and in part by the Interdisciplinary Research Program of Hebei University under Grant DXK202409. (Corresponding author: \emph{Liang Li})} \thanks{Zhou He and Ning Li are with the School of Electrical and Control Engineering, Shaanxi University of Science and Technology, Xi'an 710021, China. (email: hezhou@sust.edu.cn; 230612026@sust.edu.cn)} \thanks{Ruotian Liu is with the Department of Electrical and Information Engineering, Polytechnic University of Bari, 70125 Bari, Italy (email: ruotian.liu@poliba.it)} \thanks{Liang Li is with the School of Artificial Intelligence and Automation, Wuhan University of Science and Technology, Wuhan 430081, China (e-mail: liangli@wust.edu.cn)} \thanks{ Carla Seatzu is with the Department of Electrical and Electronic Engineering, University of Cagliari, 09124 Cagliari, Italy (e-mail: carla.seatzu@unica.it).}}

\maketitle
\bstctlcite{BSTcontrol}
\thispagestyle{empty}
\pagestyle{empty}

\begin{abstract}
%Scheduling plays a critical role in enhancing the performance of flexible manufacturing systems. However, existing methods encounter significant challenges in achieving real-time responsiveness when the system is subject to dynamic events. 

This paper investigates dynamic scheduling for flexible manufacturing systems (FMSs) subject to dynamic events, such as new order arrivals, temporary order cancellations, and machine failures. Traditional methods often face significant challenges in achieving real-time responsiveness under such conditions. To address this issue, the scheduling problem is formulated as a Markov decision process (MDP) with timed Petri nets, where the future evolution of the system depends exclusively on the current marking and the subsequently executed transitions, independent of historical trajectories. The state space and action space of the MDP are constructed using the notion of basis reachability graph (a compact state space representation) of Petri nets to alleviate the state explosion problem, thereby accelerating model training convergence. Meanwhile, a hierarchical dense reward function is constructed by integrating stepwise guidance with terminal evaluation. Then, a multi-agent proximal policy optimization algorithm is employed for model training under the centralized training and decentralized execution paradigm to improve scheduling efficiency. Numerical experiments are conducted involving typical dynamic events, and the results demonstrate that the proposed method can effectively handle dynamic events and achieve superior scheduling performance compared with conventional approaches.

\par
\vspace{\baselineskip}
\textit{Note to Practitioners}---Flexible manufacturing systems are typical discrete event systems characterized by dynamism, concurrency, and resource sharing, supporting multi-variety, small-batch production. Dynamic scheduling plays a critical role in enhancing FMS performance under real-time disturbances and uncertainties such as order arrivals, cancellations, and machine failures. Timed Petri nets are effective modeling and analyzing tools for FMS scheduling, yet traditional state space search methods become computationally expensive in dynamic environments. In this work, we combine a multi-agent deep reinforcement learning algorithm with timed Petri nets to enable fast, adaptive scheduling decisions. A state-space reduction method is introduced to accelerate both the learning process and online schedule generation.  Experiments on practical benchmark systems show that our method significantly reduces the makespan compared to common rule-based approaches, while maintaining real-time responsiveness for dynamic events. This leads to shorter production cycles, higher resource utilization, and robust handling of dynamic events.  The proposed approach offers a scalable and efficient solution for intelligent manufacturing systems operating under dynamic conditions, and can be implemented with moderate computational resources.
\end{abstract}

%Flexible manufacturing systems are typical discrete event systems characterized by dynamism, concurrency, and resource sharing, and are capable of supporting multi-variety, small-batch production. Dynamic scheduling plays a critical role in enhancing the performance of flexible manufacturing systems under real-time disturbances and uncertainty. Timed Petri nets and their associated state space search algorithms provide effective tools for addressing FMS scheduling problems. In this work, a multi-agent deep reinforcement learning algorithm combined with timed Petri nets is developed for FMS scheduling with dynamic event, such as order arrivals, temporary order cancellations, and machine failures. Based on a state space reduction method, the learning efficiency and online scheduling is accelerated , which further elevates the overall optimization performance of the scheduling algorithm in terms of makespan and resource utilization. Empirical validation on practical benchmark systems shows that the proposed method significantly reduces makespan compared with traditional rule‑based approaches, while maintaining effective real‑time responsiveness for dynamic sheduling. This makes it a scalable and efficient solution for intelligent manufacturing systems under dynamic conditions.


\IEEEpeerreviewmaketitle
\section{Introduction} \label{1}
\IEEEPARstart{W} {ith} the advancement of Industry 4.0 and intelligent manufacturing, flexible manufacturing systems (FMSs) have been widely adopted in industries such as aerospace and automotive manufacturing due to their capability to process multiple part types simultaneously through alternative process routes \cite{Rossit2018}. In FMSs, scheduling tasks involve determining the operation timing and resource allocation for each process under constraints such as process precedence, shared resources, and route flexibility, to optimize key performance indicators, including makespan and cost \cite{Huang2023Survey}. However, increasing product customization and market uncertainty have introduced frequent dynamic events into the production process, such as machine failures and order insertions or cancellations, resulting in an exponential growth of the system state space \cite{Derigent2020}.

FMS scheduling is a classical NP-hard combinatorial optimization problem \cite{Fanti2016}. A wide range of methods has been developed to address this challenge. A fundamental issue in scheduling research is the construction of system models that can accurately capture system behavior while maintaining computational tractability. Leveraging a unified graphical and mathematical formalism, Petri nets (PNs) can succinctly represent key manufacturing behaviors, including sequencing, concurrency, synchronization, and conflict. Moreover, their token-based dynamic properties enable natural tracking of the evolution of an FMS \cite{He2021, He2017, He2018, fanti2026survey}. In addition, the structural scale of PNs grows linearly with the number of system components, resulting in high modeling efficiency\cite{Mazi2022}. Owing to these advantages, scheduling research based on PNs has evolved into three main directions: metaheuristic, tree search, and machine learning.

Metaheuristic methods explore the solution space through swarm intelligence mechanisms and demonstrate strong flexibility in handling large scaled problems \cite{Prakash2011}. Xing \textit{et al.} \cite{KeYiXing2012} incorporated an optimal polynomial complexity deadlock avoidance strategy into a genetic algorithm and repaired infeasible chromosomes using a one-step look-ahead method, thereby demonstrating the effectiveness of integrating PN-based deadlock control with evolutionary algorithms. Han \textit{et al.} \cite{Han2015} first applied particle swarm optimization to deadlock free scheduling in deadlock prone FMSs, significantly enhancing optimization performance and robustness through particle normalization and simulated annealing based local search. Luo \textit{et al.} \cite{Luo2020} proposed a perturbation strategy based on estimation of distribution and a Pareto dominance acceptance criterion, which improves the global exploration capability of local search in FMS scheduling. Although metaheuristic methods alleviate the challenge of search complexity to some extent, they still suffer from issues such as unstable convergence speed and inconsistent solution quality.

Lee \textit{et al.} \cite{DooYongLee1994} were the first to introduce the A* search algorithm based on PN reachability graphs (RG) to address the scheduling problem of FMSs. Subsequently, extensive research has been conducted on the design of admissible heuristic functions \cite{Yuan2025}. Huang \textit{et al.} \cite{Huang2022}  proposed heuristic functions capable of handling token remaining time, weighted arcs, and multiple resource copies. Yi and Luo \cite{Yi2025} introduced the concept of artificial potential fields into heuristic design. However, these methods suffer from the state space explosion problem, wherein the number of states in the RG grows exponentially with system scale, resulting in a sharp increase in memory consumption and computational time. To mitigate this issue, researchers have introduced beam search and its variants. Luo \textit{et al.} \cite{JianChaoLuo2015} proposed a dynamic window search algorithm to control computational complexity by limiting the size of the search window. As an important truncated search strategy, beam search constrains space overhead by retaining only a limited number of promising candidate nodes at each layer and has demonstrated strong performance in large scaled FMSs\cite{Cherif2021}. The iterative hybrid filtered beam search algorithm proposed in \cite{Meja2017} employs a dual-filtering mechanism and an iterative deepening strategy to effectively alleviate state explosion in PN reachability graphs. He \textit{et al.} \cite{He2026} utilized the basis reachability graph (BRG) to compress the state space and combined it with beam search, significantly improving the efficiency of solving large scaled problems.

\begin{figure*}[!htbp]
  \centering
  \includegraphics[scale=1.1]{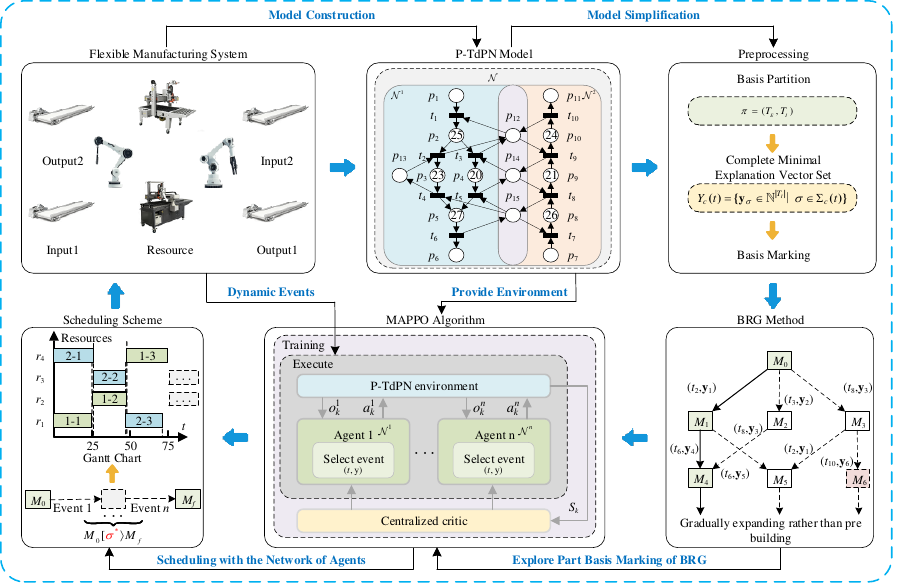}
  %[width=\textwidth, trim=0 0 3cm 0, clip]{myfigure/Flow.pdf}
  \caption{Flow chart of the proposed scheduling method.}
  \label{FlowChart}
\end{figure*}

With the increasing uncertainty in manufacturing environments, scheduling problems have gradually evolved from static to dynamic settings. In this context, reinforcement learning (RL) has attracted considerable attention due to its trial-and-error learning paradigm. Drakaki and Tzionas  \cite{Drakaki2017} employed timed colored PNs to construct an RL environment and ensured system liveness and deadlock-free properties through state-space analysis. Lee and Kim \cite{Lee2021} incorporated the idle and waiting times of bottleneck machines as penalty signals, enabling the learned policy to significantly outperform conventional scheduling rules under uncertain processing times. The method proposed in \cite{Yao2024} utilizes sparse Q-tables to reduce storage costs while improving scheduling performance and maintaining efficient online decision making. However, these methods rely on Q-tables to represent the value function, which introduces a fundamental scalability limitation when applied to complex systems with large state spaces.

Deep reinforcement learning (DRL) leverages deep neural networks to approximate value functions or policy functions, thereby overcoming the dimensional limitations of tabular methods and providing new opportunities for solving large scaled and complex scheduling problems \cite{Lassoued2024, Li2024, Hu2020, Li2022, liu2026deep}. Hu \textit{et al.} \cite{Hu2020} proposed a PN convolution layer that utilizes the input–output matrix of a PN for feature propagation, enabling deep Q-networks to effectively address dynamic scheduling for FMSs with stochastic product arrivals. Luo \textit{et al.} \cite{Luo2024} further developed a graph convolutional network that preserves the topology of the PN, allowing the model to adapt to structural changes caused by new orders and equipment failures and to achieve real time scheduling in automated manufacturing systems. Zhang \textit{et al.} \cite{Zhang2022} proposed a cloud–edge collaborative distributed real time scheduling framework for cloud manufacturing scenarios, modeling the scheduling problem as a semi-Markov decision process and solving it using a dueling deep Q-network, achieving superior performance in terms of weighted tardiness, average flow time, and other metrics.

Although existing studies have made significant progress in FMS scheduling, the integration of learning-based methods remains at an exploratory stage, particularly in dynamic environments, where scheduling approaches require further investigation. By decoupling an FMS into multiple interrelated production units and constructing corresponding scheduling agents for each unit, the complexity of individual decision spaces can be reduced, while global performance optimization can be achieved through collaborative learning among multiple agents. In this paper, a dynamic scheduling method that integrates PNs with multi-agent DRL is proposed to minimize the makespan in FMSs. The overall workflow of the proposed BRG-based MAPPO scheduling framework is shown in Fig~\ref{FlowChart}, and the main contributions are summarized as follows.

\begin{enumerate}
\item A dynamic scheduling framework based on the place-timed Petri net (P-TdPN) is developed for FMSs, where the system structure is decomposed to assign independent scheduling agents to different production lines.
\item The FMS scheduling problem is formulated as an MDP using P-TdPN, and the state space is represented using a compact structure of the state space (i.e., BRG)  and augmented with composite local observations, while the action space is constructed offline using a complete minimal explanation vector set, thereby eliminating the need for online recomputation of minimal explanations.
\item A BRG-guided multi-agent proximal policy optimization (MAPPO) scheme is proposed, in which the BRG provides a compact representation of the state space that preserves the fundamental structural and behavioral properties of the original system, while accelerating the convergence speed of model training, boosting online scheduling efficiency, and enhancing the real-time performance of dynamic scheduling response.
\item Extensive experiments under representative disturbances, including new order arrivals, temporary order cancellations, and machine failures, demonstrate that the proposed method consistently outperforms rule-based baselines.
\end{enumerate}

This paper is organized as follows. Section~\ref{Preliminary} presents the necessary preliminaries, including the theoretical foundations required for this paper. Section~\ref{4} describes the MAPPO-based scheduling algorithm built upon the BRG. The simulation results and comparative evaluations on benchmark systems are discussed in Section~\ref{5}. Finally, Section~\ref{conclusion} concludes the paper and outlines directions for future work.

\section{System Scheduling with Timed PNs}\label{Preliminary}
\subsection{Petri Net}
A PN is defined as a four-tuple $N = (P, T, Pre, Post)$ \cite{Ezpeleta1995}, in which $P = \{p_1, p_2, \ldots, p_m\}$ constitutes a finite set of $m$ places and $T = \{t_1, t_2, \ldots, t_n\}$ constitutes a finite set of $n$ transitions, satisfying $P \cap T = \emptyset$. The two mappings $Pre, Post: P \times T \rightarrow \mathbb{N}$ serve as the pre-incidence and post-incidence functions, respectively, which encode the weighted arc connections within the net, where $\mathbb{N} = \{0, 1, 2, \ldots\}$ stands for the set of natural numbers. The incidence matrix is defined as $C = Post - Pre \in \mathbb{Z}^{m\times n}$, where $\mathbb{Z}$ is the integer set.

For a transition $t \in T$, ${}^{\bullet}t = \{p \in P \mid Pre(p, t) > 0\}$ identifies its input places and $t^{\bullet} = \{p \in P \mid Post(p, t) > 0\}$ identifies its output places. Symmetric definitions apply to every place $p \in P$: ${}^{\bullet}p$ and $p^{\bullet}$ collect the input and output transitions of $p$, respectively.

A marking $M: P \rightarrow \mathbb{N}$ assigns a non-negative integer to each place, indicating the number of tokens residing in it. It is conventionally written as a column vector $M \in \mathbb{N}^m$, with the component $M(p)$ recording the token count in place $p$. A PN equipped with an initial marking $M_0$ constitutes a \textit{PN system}, denoted as $\langle N, M_0\rangle$.

A transition $t \in T$ is enabled at marking $M$, indicated as $M[t\rangle$, if  $M(p) \geq Pre(p, t)$ holds for all $p \in {}^{\bullet}t$. Once enabled, $t$ can fire and produce a successor marking
\begin{equation}
\label{eq:pn_fire}
M' = M + C(\cdot, t),
\end{equation}
denoted by $M[t\rangle M'$. A transition sequence is a finite ordered sequence $\sigma = t_1 t_2 \cdots t_k$ with $t_i \in T$. The set of all finite sequences of
transitions over $T$ is denoted by $T^*$. Then $M[\sigma\rangle M'$ denotes
that a transition sequence $\sigma \in T^*$ is enabled at $M$ and $M'$ is reachable from $M$ by firing $\sigma$. Associated with each $\sigma \in T^*$ is a firing count vector $\mathbf{y}_\sigma$, and $\mathbf{y}_\sigma(t)$ records the number of occurrences of transition $t$ within $\sigma$. The set of all markings that can be reached from $M_0$ is denoted by $\mathcal{R}(N, M_0)$.

\subsection{Timed Petri Net for Scheduling}
In FMSs, multiple production lines typically execute their respective process routes in parallel while competing for a set of shared manufacturing resources. For each processing route, a PN submodel can be constructed to accurately describe its dynamic behavior, including state transitions, event triggering, and resource allocation. These submodels can independently represent processes such as workpiece processing, resource utilization, and material transportation. Subsequently, the submodels are integrated through shared places to form a complete system model. However, if the entire system is modeled as a single, monolithic scheduling process, the state transition structure becomes increasingly complex with the growth in the number of jobs, machines, and routing schemes, which hinders scalable decision-making. The resulting place-timed Petri net (P-TdPN) \cite{Zhou2023} is formally defined as follows.

\begin{definition} \label{Def1}
A P-TdPN for the FMS scheduling is $\mathcal{N} = (P, T, Pre, Post, D)$, where the place set is partitioned as $P = P_S \cup P_O \cup P_E \cup P_R$ with $P_S$, $P_O$, $P_E$, and $P_R$ denoting the sets of start places, operation places, end places, and resource places, respectively, and $D: P_O \rightarrow \mathbb{N}$ is a delay function that assigns a deterministic processing time to each operation place.
In particular, we denote by $\mathcal{N}^i = (P_S^i \cup P_O^i \cup P_E^i \cup P_R^i, T^i, Pre^i, Post^i, D^i)$ the subnet corresponding to the $i$-th production line, where $P^i$, $T^i$, $Pre^i$, $Post^i$, and $D^i$ are the restrictions of $P$, $T$, $Pre$, $Post$, and $D$ to this subnet. Here, $i \in \mathbb{J}$, with $\mathbb{J} = \{1, 2, \dots, J\}$ denoting the index set of production lines. These subnets are integrated through shared resource places. The following properties are satisfied:

\begin{enumerate}  
    \item $P_S^i \cap P_O^i = P_E^i \cap P_O^i = \emptyset$ and $(P_S^i \cup P_O^i \cup P_E^i) \cap P_R^i = \emptyset$; $P_S = \bigcup_{i=1}^{k} P_S^i$, $P_O = \bigcup_{i=1}^{k} P_O^i$, $P_E = \bigcup_{i=1}^{k} P_E^i$, $P_R = \bigcup_{i=1}^{k} P_R^i$, $T = \bigcup_{i=1}^{k} T^i$.
    
    \item For all $i, j \in \mathbb{J}$ with $i \neq j$, it holds that $(P_S^i \cup P_O^i \cup P_E^i) \;\cap\; (P_S^j \cup P_O^j \cup P_E^j) = \emptyset$, and $T^i\cap T^j = \emptyset$.
    
    \item For all $i, j \in \mathbb{J}$ with $i \neq j$, the intersection of the place sets of $\mathcal{N}^i$ and $\mathcal{N}^j$ is contained in $P_R$, i.e., $(P^i \cap P^j) \subseteq P_R$.

    \item For every $p \in P$ and every $t \in T$, both $Pre(p,t)$ and $Post(p,t)$ belong to $\{0,1\}$.
    
    \item The delay function $D^i: P_O^i \rightarrow \mathbb{N}$ assigns a deterministic processing time to each operation place on production line $i \in \mathbb{J}$.
\end{enumerate}
\end{definition}

In Definition \ref{Def1}, item 1) specifies the structural composition of each subnet and the global model. In particular, the start, operation, and end places are mutually disjoint within each subnet and are separated from resource places, ensuring a clear distinction between process flow and resource representation; item 2) enforces that different production line subnets are independent in terms of process places and transitions; item 3) indicates that interactions among different subnets occur exclusively through resource places; item 4) restricts the arc weights of the PN to be binary, implying that each transition consumes and produces at most one token from each connected place. This assumption simplifies the model while remaining sufficient to describe typical processing and resource allocation behaviors; item 5) defines the temporal semantics of the model by assigning a deterministic processing time to each operation place. Specifically, upon entering an operation place $p \in P_O^i$, a token remains temporally unavailable for a duration specified by $D^i(p)$, representing an ongoing manufacturing operation. Only after this deterministic delay elapses does the token become available to enable subsequent downstream transitions.

\begin{figure}[!htbp]
  \centering
  \includegraphics[scale=0.7]{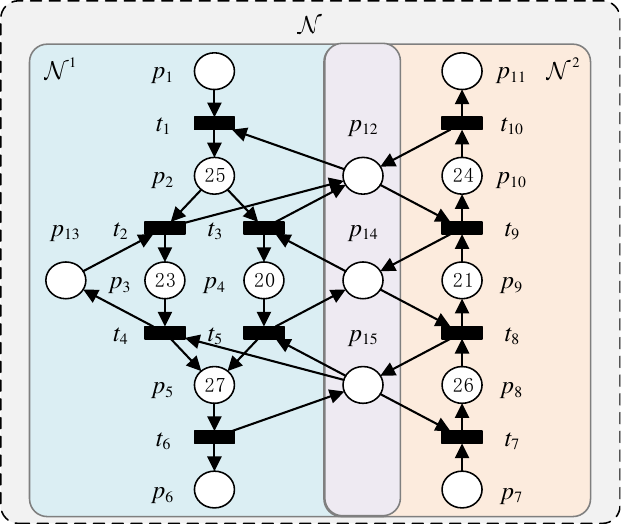}
  \caption{The P-TdPN of the FMS in Example 1.}
  \label{PTPN_Indus}
\end{figure}

\begin{example} \label{example1}
Let us consider an industrial example adapted from \cite{Han2015}, whose layout is illustrated in Fig.~\ref{Indus}. The FMS consists of four types of resources, denoted as $R_1$–$R_4$, each with a single unit capacity. It also includes two loading buffers, $I_1$ and $I_2$, and two unloading buffers, $O_1$ and $O_2$. Two types of jobs, $L_1$ and $L_2$, are processed collaboratively by resources. The production lines are defined as follows, where the processing time of each operation is given in parentheses, 
$L_1: I_1 \rightarrow R_1(25) \rightarrow R_2(23) \rightarrow R_4(27) \rightarrow O_1$, or $I_1 \rightarrow R_1(25) \rightarrow R_3(20) \rightarrow R_4(27) \rightarrow O_1$;
$L_2: I_2 \rightarrow R_4(26) \rightarrow R_3(21) \rightarrow R_1(24) \rightarrow O_2$.

\begin{figure}[!htbp]
  \centering
  \includegraphics[scale=0.65]{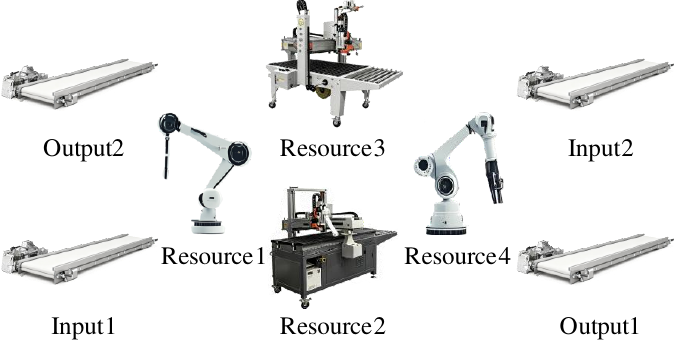}
  \caption{An industrial FMS.}
  \label{Indus}
\end{figure}

The P-TdPN of the FMS is illustrated in Fig.~\ref{PTPN_Indus}. According to the definition, the system is decomposed into two production-line subnets, denoted as $\mathcal{N}^1 = (P_S^1 \cup P_O^1 \cup P_E^1 \cup P_R^1, T^1, Pre^1, Post^1, D^1)$ and $\mathcal{N}^2 = (P_S^2 \cup P_O^2 \cup P_E^2 \cup P_R^2, T^2, Pre^2, Post^2, D^2)$. For subnet $\mathcal{N}^1$, the place sets are defined as $P_S^1 = \{p_1, p_2, p_3, p_4, p_5\}$, $P_O^1 = \{p_2, p_3, p_4, p_5\}$, $P_E^1 = \{p_6\}$, and $P_R^1 = \{p_{12}, p_{13}, p_{14}, p_{15}\}$, which correspond to resources $R_1$–$R_4$, respectively. The transition set is given by $T^1 = \{t_1, t_2, t_3, t_4, t_5, t_6\}$. The subnet $\mathcal{N}^2$ can be constructed analogously. These subnets are interconnected through shared resource places, i.e., $P_R^1 \cap P_R^2 = \{p_{12}, p_{14}, p_{15}\}$. The overall P-TdPN model for FMS scheduling is defined as $\mathcal{N} = (P, T, Pre, Post, D)$, where $P = P_S^1 \cup P_O^1 \cup P_E^1 \cup P_R^1 \cup P_S^2 \cup P_O^2 \cup P_E^2 \cup P_R^2$ and $T = T^1 \cup T^2$. \hfill $\square$
\end{example}

\subsection{Problem Statement}
Given a P-TdPN $\mathcal{N} = (P, T, Pre, Post, D)$ for an FMS as defined in Definition~\ref{Def1}, the dynamic scheduling problem can be formally stated as follows. The system starts from an initial marking $M_0$, in which tokens in start places $P_S$ represent jobs to be processed, and tokens in resource places $P_R$ indicate the availability of manufacturing resources. A \textit{final marking} $M_f$ is reached when all jobs have been completed, characterized by $M_f(p) = 0$ for all $p \in P_S \cup P_O$ and $M_f(p) = M_0(p)$ for all $p \in P_E \cup P_R$, indicating that all tokens have migrated from start places to end places through their respective process routes, while all resources have been released.

During the production process, the system is subject to three types of dynamic events whose occurrence produces a change in the current marking of the P-TdPN for the FMS.
\begin{itemize}
    \item \textit{New order arrivals}: Additional jobs arrive dynamically at unpredictable times, requiring insertion into the current schedule. This is modeled by adding tokens to the corresponding start places $P_S^i$ of the affected production line subnet $\mathcal{N}^i$.
    
    \item \textit{Order cancellations}: Existing jobs may be cancelled due to changes in customer demand. This event is handled by removing tokens from the relevant places in $P_S^i \cup P_O^i$ and releasing the associated resources.
    
    \item \textit{Machine failures}: Resources may become temporarily unavailable due to unexpected breakdowns. This is represented by removing tokens from the affected resource places in $P_R$ for the duration of the failure, and restoring them upon repair completion.
\end{itemize}
%Note that the 

%\begin{problem}
%Let $\sigma \in T^*$ denote a complete transition sequence satisfying $M_0[\sigma\rangle M_f$, and let $\mathcal{C}(\sigma)$ denote the makespan associated with $\sigma$, and
%\begin{equation}
%F_{\max}(\sigma)=\max_{j \in \mathbb{N}_J}\left\{\tau_j\right\}
%\end{equation}
%is the makespan, with $\tau_j^{\mathrm{fin}}$ being the completion time of the last operation of job $j$. The terminal marking $M_f$ is reached when all jobs have been completed and every resource token has been returned to its original availability..
%\end{problem}

\textit{Problem 1} (Dynamic Scheduling Problem): Based on the P-TdPN $\mathcal{N}$, the dynamic scheduling of FMSs with $J$ production lines under dynamic events (i.e., new order arrivals, order cancellations, machine failures) can be formulated as determining a feasible firing sequence of transitions $\sigma^\ast$ that drives the system from the initial marking $M_0$ to the final marking $M_f$, i.e., $M_0[\sigma^\ast\rangle M_f$,  while minimizing the makespan $g(M_f, \sigma^\ast)$. The terminal marking $M_f$ is reached when all jobs have been completed and every resource token has been returned to its original availability. $\hfill$ $\blacksquare$

The challenge lies in achieving real-time responsiveness: whenever a dynamic event occurs, the scheduling policy must rapidly generate an updated decision without exhaustive re-computation of the entire state space. Traditional optimization methods that rely on enumeration of $\mathcal{R}(N, M_0)$ become computationally prohibitive in dynamic settings, particularly as the system scale increases. This motivates the development of learning-based approaches that can adapt online to evolving system states while maintaining near-optimal performance.

\subsection{Basis Reachability Graph}
The BRG provides a compact representation of the state space \cite{Ma2017}. Similar to the conventional RG, it can be regarded as a deterministic structure. The BRG contains all basis markings reachable from the initial marking $M_0$. Before presenting the construction procedure of the BRG, several related concepts are introduced. For a more detailed discussion, readers are referred to \cite{Ma2017} and \cite{Ma2022}.

\begin{definition}\cite{Ma2017}
Given a PN $N=(P,T,Pre,Post)$, a pair $\pi=(T_E,T_I)$ of the transition set $T$, with $T=T_E \uplus T_I$, is called a \emph{basis partition} if the $T_I$-induced subnet is acyclic. The sets $T_E$ and $T_I$ are referred to as \emph{explicit transitions} and \emph{implicit transitions}. 
%Let $C_I\in\mathbb{Z}^{|P|\times |T_I|}$ denote the incidence matrix associated with the implicit subnet.
\end{definition}

\begin{definition} \label{Ymin}\cite{Ma2017}
Given a PN $N$, a reachable marking $M$, a basis partition $\pi=(T_E,T_I)$ and an explicit transition $t\in T_E$, we define
\begin{itemize}
    \item $\Sigma(M,t)=\{\sigma\in T_I^{*}\mid M[\sigma t\rangle\}$ as the set of \textit{explanations} of $t$ at $M$;
    \item $Y(M, t) = \{\mathbf{y}_{\sigma} \in \mathbb{N}^{|T_I|} \mid \sigma \in \Sigma(M, t)\}$ as the set of \textit{explanation vectors};
    \item $\Sigma_{min}(M, t) = \{\sigma \in \Sigma(M, t) \mid \nexists \sigma' \in \Sigma(M, t), \mathbf{y}_{\sigma'} \lneq \mathbf{y}_{\sigma}\}$ as the set of \textit{minimal explanations};
    \item $Y_{min}(M, t) = \{\mathbf{y}_{\sigma} \in \mathbb{N}^{|T_I|} \mid \sigma \in \Sigma_{min}(M, t)\}$ as the set of \emph{minimal explanation vectors}.
\end{itemize}
\end{definition}

Intuitively, an explanation $\sigma \in \Sigma(M, t)$ represents a sequence of implicit transition firings that must occur before the observed transition $t$ can fire at marking~$M$. Since multiple implicit sequences may enable~$t$, the set $\Sigma(M, t)$ generally contains several explanations. Note that different sequences $\sigma, \sigma' \in \Sigma(M, t)$ may share the same vector $\mathbf{y}_{\sigma} = \mathbf{y}_{\sigma'}$. Hence, in general, $|Y(M,t)| \leq |\Sigma(M,t)|$. $Y_{min}(M, t)$ denotes the set of all minimal elements of $Y(M, t)$.

\begin{definition}\cite{Ma2017}
Given a PN system  $\langle N, M_0\rangle$ and a basis partition $\pi = (T_E, T_I)$, the set of \emph{basis markings} of the system is defined as  $\mathcal{M}(N, M_0, \pi)\subseteq \mathcal{R}(N, M_0)$, such that:
\begin{equation}
\begin{aligned}
\mathcal{M}(N, M_0, \pi) = \{\{M_0, M'\} \mid & M' = M_0 + C_I \cdot \mathbf{y} + C(\cdot, t), \\ & \mathbf{y} \in Y_{min}(M, t), t \in T_E\},
\end{aligned}
\end{equation}
where $C_I$ denotes the incidence matrix $C$ restricted to $P \times T_I$. A new basis marking is obtained by firing an explicit transition $t \in T_E$ together with its corresponding minimal explanation.
\end{definition}

\begin{definition}\cite{Ma2017}
\label{BRG}
Given a PN system  $\langle N, M_0\rangle$ and a basis partition $\pi = (T_E, T_I)$, the corresponding BRG is defined as $\mathcal{B} = (\mathcal{M}(N, M_0, \pi), T_r, \Delta)$, where  
\begin{itemize}
    \item the set of states $\mathcal{M}_B$ consists of all basis markings;
    \item the event set $T_r$ is composed of pairs $(t, \mathbf{y}) \in T_E \times \mathbb{N}^{|T_I|}$;
    \item $\Delta \subseteq \mathcal{M}(N, M_0, \pi) \times T_r \times \mathcal{M}(N, M_0, \pi)$ denotes the transition relation defined by  
    $\Delta = \{(M, (t, \mathbf{y}), M') \mid \mathbf{y} \in Y_{min}(M, t), t \in T_E, M' = M + C_I \cdot \mathbf{y} + C(\cdot, t)\}$.
\end{itemize}
\end{definition}

From Definition \ref{BRG}, it follows that a PN  may correspond to different BRGs under distinct basis partitions $\pi = (T_E, T_I)$. At present, a precise quantitative relationship between the size of a BRG and the selection of basis partition $\pi$ has not been established. In this paper, the partitioning method proposed in \cite{Ma2017} is adopted to construct an appropriate basis partition, aiming to reduce the scale of the resulting BRG.

\begin{proposition}
\label{prop:1}
\cite{Ma2017}
Consider a PN and its corresponding BRG $\mathcal{B} = (\mathcal{M}(N, M_0, \pi), T_r, \Delta)$ under the basis partition $\pi = (T_E, T_I)$. The following statements are equivalent:
\begin{itemize}
    \item There exist a marking $M$ and a transition sequence 
    $\sigma = \sigma_1 t_1 \cdots \sigma_n t_n \sigma_{n+1}$, where $\sigma_i \in T_I^*$ and $t_i \in T_E$ for all $i = 1,...,n$, such that $M_0[\sigma\rangle M$;
    
    \item There exists a path in the BRG of the form
    \begin{equation*}
        M_0 \xrightarrow{(t_1, \mathbf{y}_1)} M_{1} 
        \xrightarrow{(t_2, \mathbf{y}_2)} \cdots 
        \xrightarrow{(t_n, \mathbf{y}_n)} M_{n},
    \end{equation*}
    where $\mathbf{y}_i \in Y_{\min}(M_{i-1}, t_i)$ for all $i = 1, \dots, n$, and the marking $M$ satisfies $M \in \{ M' \mid M_n[\sigma'\rangle M',\ \sigma' \in T_I^* \}$.
\end{itemize}
\end{proposition}

In general, obtaining a transition sequence $\sigma^{\ast}$ that satisfies $M_0[\sigma^{\ast} \rangle M_f$ requires constructing the complete set of RG of a PN. This process leads to substantial memory consumption and increases the computational burden of scheduling. Based on Proposition~\ref{prop:1}, any transition sequence $\sigma$ consisting of both implicit and explicit transitions can be equivalently represented through a sequence derived from the BRG, where only implicit transitions are involved in the expansion process. As a result, the BRG is significantly more compact than the RG, which helps reduce storage requirements and enhances computational efficiency. {Scheduling is performed by incrementally exploring the BRG of the P-TdPN rather than constructing the complete BRG in advance. During this process, the BRG is used solely to represent the state information of the system, while temporal information is computed using the idle interval scheduling method described in Section~III-C. This is consistent with traditional RG-based scheduling methods, in which the RG serves only as a state representation and does not explicitly encode temporal information \cite{Huang2022}. Consequently, the exploration of the BRG for a P-TdPN shares the same fundamental structure as the exploration of the BRG for an ordinary PN.}

\begin{figure}[!htbp]
  \centering
  \includegraphics[scale=0.5]{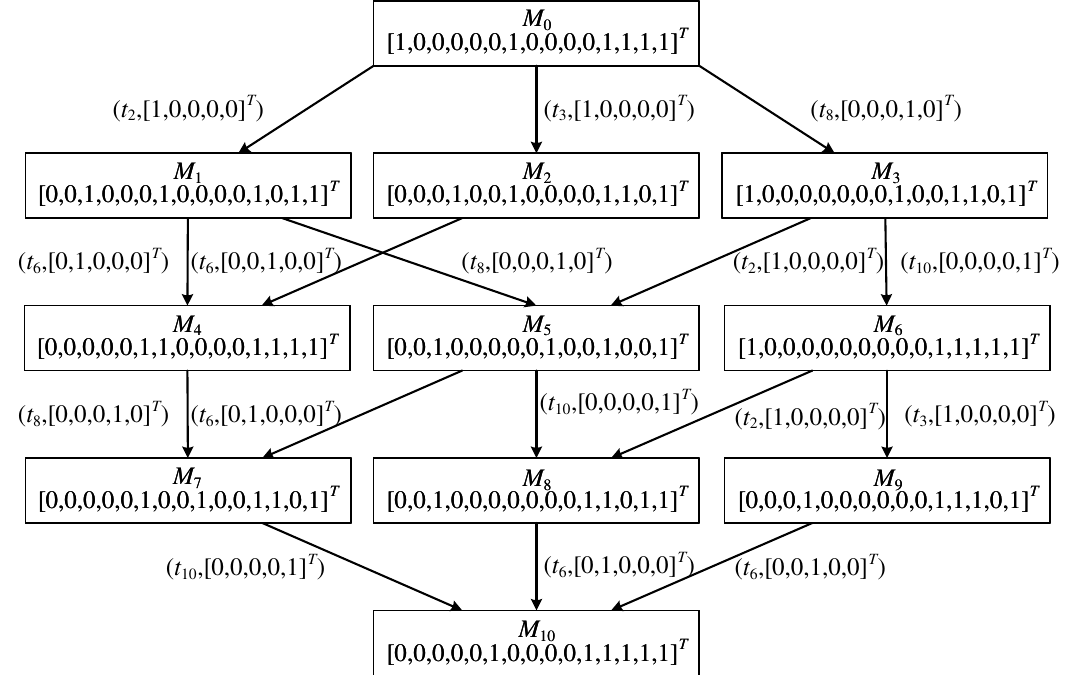}
  \caption{The BRG $\mathcal{B}$ of the P-TdPN in Fig. \ref{PTPN_Indus}.}
  \label{example_BRG}
\end{figure}

\begin{example}
\label{example2}
(\textit{Example \ref{example1} continued}) Consider the P-TdPN in Fig. \ref{PTPN_Indus} with an initial marking $M_0 = [1,0,0,0,0,0,1,0,0,0,0,1,1,1,1]^T$. The RG of the net has 26 reachable markings. However, under a basis partition $\pi = (T_E, T_I)$ with $T_E = \{t_{2}, t_{3}, t_{6}, t_{8}, t_{10}\}$ and $T_I = T \setminus T_E$, where the partition is constructed using the method proposed in \cite{Ma2017} to reduce the scale of the resulting BRG, the resulting BRG has only 11 basis markings, as shown in Fig. \ref{example_BRG}. \hfill $\square$
\end{example}

\section{Deep Reinforcement Learning for Scheduling Based on  BRG}\label{4}

In this section, the dynamic scheduling problem of FMSs is formulated as an MDP, in which the P-TdPN serves as the interactive environment for the agents. The proposed scheduling framework adopts MAPPO \cite{yu2022surprising}, a multi‑agent deep reinforcement learning algorithm, to learn scheduling policies for each production line. {MAPPO is suitable for this problem due to its ability to handle cooperative multi-agent decision making under shared resource constraints, and involves cooperative decision making in a discrete action space. In addition, the centralized-training-and-decentralized-execution paradigm facilitates coordination among production lines sharing limited resources, while the PPO-based clipped update mechanism provides stable learning performance.} As the execution entities of the MAPPO algorithm, the agents are responsible for observing the environment state, generating scheduling decisions, and executing corresponding actions to accomplish the scheduling task.

An MDP is typically defined as a tuple $(S, A, \mathcal{P}, r, \zeta)$, where $S$ denotes the state space, $A$ represents the action space, $\mathcal{P}(s' \mid s, a)$ is the state transition probability, $r(s, a)$ is the immediate reward function, and $\zeta \in [0,1)$ is the discount factor. At each step, an agent selects an action based on the current state, and the environment transitions to a new state according to $\mathcal{P}$ while yielding a reward. The objective is to learn an optimal policy that maximizes the expected cumulative discounted reward over time. We define the MDP over the states, actions, and rewards in the proposed framework as follows.

\subsection{State Representation}
The proposed scheduling framework is built under the centralized training and decentralized execution paradigm of MAPPO. Each agent is assigned to one production line. The state representation should satisfy two requirements simultaneously: i) the actor of each agent should rely only on the local information of its own production-line subnet, and ii) the centralized critic should exploit the joint information of all production lines so as to capture the coupling induced by shared resources. The P-TdPN model for FMS scheduling in Definition~\ref{Def1} provides a natural basis for such a design, as the marking of a subnet directly reflects its local job progress and resource occupancy.

Let $M_k \in \mathcal{M}(N, M_0, \pi)$ denote the global marking of the P-TdPN at step $k$. For production line $i \in \mathbb{J}$, we have $P^{i}=P_S^{i}\cup P_O^{i}\cup P_E^{i}\cup P_R^{i}$, and let $R^{i}$ be the set of resource types involved in subnet $\mathcal{N}^{i}$. The local observation of agent $i$ at step $k$ is defined as
\begin{equation}
\label{eq:local_obs}
o_k^{i} = [e_k^{i}, c_k^{i}, \ell_k^{i}, u_k^{i}]
\end{equation}
where $e_k^{i}$ denotes the token distribution over the places of subnet $\mathcal{N}^i$ (local marking), $c_k^{i}$ is the residual processing time, $\ell_k^{i}$ is the residual load, and $u_k^{i}$ is the cumulative resource usage of each resource type. A formal definition of them is provided in the following. The four-component observation vector is designed to provide complementary views of the scheduling state: $e^i_k$ captures the instantaneous job progress, $c^i_k$ and $\ell^i_k$ provide forward-looking estimates of the remaining workload at the system and resource levels respectively, while $u^i_k$ records the historical resource utilisation to balance load across machines. These four components are detailed below.

\textit{Local marking}:
The marking is the most direct state descriptor provided by the P-TdPN. The local marking of subnet $\mathcal{N}^{i}$ is defined as the projection of the global marking onto the places contained in $P^{i}$:
\begin{equation}
\label{eq:local_marking}
e_k^{i}=\big[M_k(p)\big]_{p\in P^{i}}.
\end{equation}
This observation records the distribution of tokens over the start, operation, end, and resource places of production line $i$. If a resource place is shared by multiple subnets, its marking is visible to every agent whose subnet contains that place.

\textit{Residual processing time}:
To endow the agent with a forward looking awareness of scheduling progress, we introduce a feature that estimates the minimum total processing time required to complete all unfinished operations. For each place $p \in P^{i} \setminus P_R^{i}$, let $\mathbf{h}^{i}(p) \geq 0$ denote the minimum cumulative processing delay from place $p$ to an end place in $P_E^{i}$, computed as the sum of the deterministic processing times $D^i(p')$ assigned by the delay function to all operation places $p' \in P_O^i$ along the shortest path from $p$ to $P_E^i$ in the subnet topology. This computation assumes that each token departs an operation place immediately upon completing its prescribed sojourn of $D^i(p')$ time units, thereby neglecting additional waiting times induced by resource contention or blocking. By definition, $\mathbf{h}^{i}(p) = 0$ for every place $p \in P_E^{i}$, and $\mathbf{h}^{i}(p) > 0$ only for every place $p \in P_S^{i} \cup P_O^{i}$. The residual processing time of agent $i$ at step $k$ is then defined as
\begin{equation}
\label{eq:rho}
c_k^{i} = \big[ M_k(p) \cdot \mathbf{h}^{i}(p) \big]_{p \in P^{i} \setminus P_R^{i}}.
\end{equation}
Although $c_k^{i}$ ignores future blocking and resource conflicts, it provides a compact estimate of the remaining processing workload and decreases monotonically as jobs progress toward completion.

\textit{Residual load}:
In addition to global completion progress, an effective agent should capture how future workloads are distributed across resources. For each subnet $\mathcal{N}^{i}$, $\Gamma^{i}(p,r) \geq 0$ denotes the weighted resource time matrix proposed in \cite{Huang2022}, where $p \in P^{i} \setminus P_R^{i}$ and $r \in R^{i}$, representing the minimum remaining processing time on resource $r$ for a token at place $p$. The residual load is defined as
\begin{equation}
\label{eq:load}
\ell_k^{i} = \big[ \sum_{p \in P^{i} \setminus P_R^{i}} M_k(p) \cdot \Gamma^{i}(p,r) \big]_{r \in R^{i}}.
\end{equation}
This vector captures the minimum future workload on each resource type for production line $i$, highlighting potential bottlenecks and load imbalances.

\textit{Cumulative resource usage}:
The descriptor in \eqref{eq:load} reflects prospective workload only. To complement the forward-looking information, we also record how much processing effort has already been consumed by each resource type. Let $\sigma_k^{i} \in (T^{i})^{*}$ be the transition sequence that has occurred in subnet $\mathcal{N}^{i}$ up to step $k$. For each $t\in \sigma_k^{i}$, let $p_o^t \in t^\bullet \cap P^i_O$ be the corresponding operation place and $p_r^t \in t^\bullet \cap P^i_R$ be the corresponding resource place. The cumulative resource  usage is defined as
\begin{equation}
\label{eq:usage}
u_k^{i} = \big[\big(\sum_{t \in \sigma_k^{i}} \delta(t, r) \cdot D(p_o^t)\big) / M(p_r^t) \big]_{r \in R^{i}},   
\end{equation}
where $D(p_o^t)$ denotes the processing time associated with the operation place $p_o^t$. $\delta(t, r) = 1$ if the resource place associated with transition $t$ is dedicated to resource $r$, and $\delta(t, r) = 0$ otherwise. Division by $M(p^t_r)$ normalises the usage by the resource capacity, yielding a per-unit utilisation metric. Together, $\ell_k^{i}$ and $u_k^{i}$ provide complementary views of resource status: the former describes the minimum remaining demand, whereas the latter reflects the workload already processed.

Under decentralized execution, the actor of agent $i$ receives only its local observation $\mathbf{o}_k^{i}$ and outputs the corresponding local scheduling action. On the contrary, during training, all local observations are concatenated to form the joint state at step $k$, denoted as
\begin{equation}
S_k=(o_k^{1},o_k^{2},\ldots,o_k^{J}),  
\end{equation}
which is provided to the centralized critic for value evaluation. The centralised critic uses $S_k$ during training to produce more accurate value estimates, while each actor uses only its local observation $o^i_k$ during execution, enabling scalable deployment without communication between agents.

\subsection{Action Representation}
{In the reinforcement learning framework based on BRG, the agent selects an event $(t, \mathbf{y})$ at each marking as the scheduling action. The selected action drives the system to evolve from the current state to a subsequent state, and through a sequence of such decisions, the system eventually reaches the final marking, thereby completing the scheduling task.} In the conventional BRG construction, the set of minimal explanation vectors $Y_{\min}(M,t)$ must be computed incrementally for each newly visited basis marking $M \in \mathcal{M}(N, M_0, \pi)$ as the graph is expanded, since the minimal explanations of an explicit transition $t \in T_E$ are marking-dependent. While a complete BRG can theoretically be constructed offline {to obtain all events $(t, \mathbf{y})$ (action space)}, exhaustive generation is computationally intractable for large-scale FMSs due to the state explosion problem. 
To overcome this limitation, {we adopt a structural concept called the complete minimal explanation set that is independent of both the current basis marking and the initial marking \cite{Ma2017}.} The key idea is to precompute, for each explicit transition, all implicit transition sequences that may potentially enable it, using only the structure of the PN and the basis partition. These precomputed sequences serve as candidate explanations and can be reused for any given marking. This precomputation fully determines the action space before any episode begins, thereby decoupling the BRG expansion from the reinforcement learning training loop and enabling stable policy optimization over a fixed action space throughout the entire training process.

{
\begin{definition}\cite{Ma2017}
Given a PN $N$, a basis partition $\pi=(T_E,T_I)$, and an explicit transition $t \in T_E$, the \emph{complete minimal explanation set} of $t$ is defined as \[ \Sigma_c(t)=\{\sigma \in T_I^{*}\mid \exists M \in \mathbb{N}^m,\ \sigma \in \Sigma_{\min}(M,t)\}, \] which contains all sequences $\sigma$ that constitute minimal explanations of $t$ at some markings (not necessarily reachable). Correspondingly, the \emph{complete minimal explanation vector set} is defined as \[ Y_c(t) =\{\mathbf{y}_{\sigma} \in \mathbb{N}^{|T_I|}\mid \sigma \in \Sigma_c(t)\}. \]
The boundedness of $Y_c(t)$ and the corresponding computation method were established in~\cite{Ma2017}.
\end{definition}}

\begin{example}
\label{example3}
(\textit{Example \ref{example1} continued}) Consider the P-TdPN in Fig. \ref{PTPN_Indus} with a basis partition $\pi = (T_E, T_I)$, where $T_E = \{t_{2}, t_{3}, t_{6}, t_{8}, t_{10}\}$ and $T_I = T \setminus T_E$. For this P-TdPN, the set of complete minimal explanation vector set $Y_c(t)$ for each explicit transition $t \in T_E$ is shown in Table~\ref{table_YMIC}. \hfill $\square$
\end{example}

\begin{table}[!htbp]
\centering
\renewcommand{\arraystretch}{1.4}
\caption{{$Y_c(t)$} for each explicit transition in Example \ref{example3}.}
\scalebox{1}{
\begin{tabular}{cccc}
\toprule
\multicolumn{1}{c}{Transition $t$} & \multicolumn{3}{c}{$Y_c(t)$} \\ \hline
\multirow{2}{*}{$t_2$}      
                        & \{$[0, 0, 0, 0, 0]^T$, 
                        & $[0, 1, 0, 0, 0]^T$,         
                        & $[1, 0, 0, 0, 0]^T$,          \\
                        & $[1, 1, 0, 0, 0]^T$\}            \\ \midrule
\multirow{2}{*}{$t_3$}      
                        & \{$[0, 0, 0, 0, 0]^T$,            
                        & $[0, 0, 0, 0, 1]^T$,            
                        & $[0, 0, 1, 0, 0]^T$,            \\
                        & $[1, 0, 0, 0, 0]^T$,            
                        & $[1, 0, 0, 0, 1]^T$,            
                        & $[1, 0, 1, 0, 0]^T$\}            \\ \midrule
\multirow{1}{*}{$t_6$}                       
                        & \{$[0, 0, 0, 0, 0]^T$,            
                        & $[0, 0, 1, 0, 0]^T$,           
                        & $[0, 1, 0, 0, 0]^T$\}            \\ \midrule
\multirow{2}{*}{$t_8$}      
                        & \{$[0, 0, 0, 0, 0]^T$,           
                        & $[0, 0, 0, 0, 1]^T$,            
                        & $[0, 0, 0, 1, 0]^T$,            \\
                        & $[0, 0, 0, 1, 1]^T$,            
                        & $[0, 0, 1, 0, 0]^T$,            
                        & $[0, 0, 1, 1, 0]^T$\}            \\ \midrule
$t_{10}$                       
                        & \{$[0, 0, 0, 0, 0]^T$,            
                        & $[0, 0, 0, 0, 1]^T$\}            \\ \bottomrule
\end{tabular}}
\label{table_YMIC}
\end{table}

Note that the precomputed set $Y_c(t)$ contains all theoretically feasible explanation vectors for each explicit transition $t\in T_E$. However, for a given state, not all vectors in $Y_c(t)$ are feasible. The subset of feasible vectors within $Y_c(t)$ therefore defines the feasible action space. To address this issue, an action masking mechanism is employed to dynamically filter infeasible actions at each step. For the $i$-th agent, the action space is defined as $A^i = \{a_0, a_1, \dots, a_{N_a}\}$, where each candidate action corresponds to a composite event $(t, \mathbf{y})$, and $N_a = 1 + \sum_{t \in T_E^i} |Y_c(t)|$. Here, $T_E^i$ denotes the set of explicit transitions of $\mathcal{N}^i$. Each agent is equipped with a null action, indicating that no scheduling action is executed at the current step, and the system state remains unchanged. This null operation is practically meaningful in two scenarios. First, when the system is in an intermediate state where none of the explicit transitions are enabled, waiting becomes the only feasible choice. {Second, when multiple agents simultaneously compete for the same resource, an agent may intentionally choose to wait in order to avoid resource conflicts.}

Accordingly, the action mask for the $i$-th agent is defined as $\mathbf{v}^i = (v_0, v_1, \dots, v_{N_a})$, where
\[
v_j =
\begin{cases}
1, & \text{if } a_j \text{ is feasible or is an idle action},\\
0, & \text{otherwise}.
\end{cases}
\]

In particular, if all non-idle actions violate the feasibility conditions, only the idle action is retained, forcing the agent to remain inactive at the current step. The masking operation is implemented at the level of the policy network by modifying the output logits \cite{Lei2022}. Let $\mathbf{z}^i = (z_0, z_1, \dots, z_{N_a})$ denote the raw logits produced by the policy network. The masked logits are computed as $\tilde{\mathbf{z}}^i = \mathbf{z}^i + (1 - \mathbf{v}^i) \cdot (-\infty)$. After applying the Softmax function, the resulting action probability distribution is given by
\[
\pi_{\theta_i}(a_j \mid o^i) = \frac{\exp(\tilde{z}_j)}{\sum_{l=0}^{N_a} \exp(\tilde{z}_{l})}.
\]

This mechanism ensures that infeasible actions are assigned zero probability, thereby preventing violations of the structural constraints imposed by the PN. Meanwhile, it preserves the flexibility of the policy network to optimize over the feasible action subset.

\subsection{Reward Function}
The reward function shapes agent behaviour toward the scheduling objective. A hierarchical dense reward function is designed by combining stepwise guidance with terminal evaluation. The reward for agent $i$ at step $k$ is defined as
\[
r^i_k = r_k^{dist} + r_k^{time} + r_k^{step} + r_k^{conf} + r_k^{term}.
\]

\textit{Distance reward}: 
This term provides a shaping signal that encourages the system state to approach the target marking. Let $M_k$ and $M_f$ denote the current and final markings, respectively. The distance reward is defined as $r_k^{dist} = - \| M_k - M_f \|_2$, which penalizes deviations from the target marking and promotes toward task completion.

\textit{Time cost penalty}: 
To reflect the impact of each decision on the makespan, the incremental time cost is incorporated as $r_k^{time} = g(M_{k-1}, \sigma_{k-1}) - g(M_k, \sigma_k)$, where $g(\cdot)$ denotes the accumulated makespan and is evaluated using the dynamic idle interval scheduling mechanism proposed in~\cite{He2026}. {Here, $\sigma_{k-1}$ and $\sigma_k$ denote the corresponding firing sequences from the initial marking $M_0$ to $M_{k-1}$ and $M_k$, respectively.} Specifically, for each resource place, the algorithm dynamically maintains the set of active processing intervals and identifies the corresponding idle time windows during which the resource utilization remains strictly below its maximum capacity. Upon the execution of a composite event $(t, \mathbf{y})$, each associated operation is assigned to the earliest feasible idle interval while satisfying all temporal precedence constraints. The value of $g(\cdot)$ is then obtained as the makespan of the resulting schedule. This term encourages actions that incur lower processing time and improve temporal efficiency.
 
\textit{Step penalty}: 
A penalty is imposed at each step: $r_k^{step} = -\lambda_{step}$ ($\lambda_{step} > 0$), which discourages unnecessary delays and promotes compact scheduling sequences.

\textit{Conflict penalty}: 
If a joint action leads to infeasibility (e.g., negative tokens due to resource conflicts), the action is rejected, and all agents receive a penalty: $r_k^{conf} = -\lambda_{conf}$, where $\lambda_{conf} > 0$. This shared penalty encourages coordinated behavior and prevents incompatible action combinations.

\textit{Terminal reward}: 
This reward is assigned when the system reaches the final marking or the maximum step limit:
\[
r_k^{term} =
\begin{cases}
- g(M_f, \sigma_f), & \text{if } M_k = M_f,\\
- \lambda_{fail}, & \text{otherwise},
\end{cases}
\]
where $\lambda_{fail}$ is a large positive constant, thereby aligning the optimization objective with makespan minimization.

The proposed reward structure integrates dense intermediate feedback with a final performance evaluation. The distance, time, and step components provide continuous guidance to mitigate reward sparsity, while the terminal reward ensures alignment with the global objective. The conflict penalty further enhances multi-agent coordination. Together, these components form an effective reward design that accelerates convergence and improves scheduling performance.

\subsection{MAPPO-based Scheduling Algorithm}
MAPPO adopts a centralized training and decentralized execution framework with an actor--critic architecture. It consists of multiple decentralized policy networks and a shared global value network. During training, the critic leverages global state information to provide accurate value estimation, while each actor learns a policy based on its local observation. During execution, agents act independently using only local observations without access to global information.

Each agent $i$ is equipped with an independent policy network $\pi_{\theta_i}(a \mid o^i)$, which maps local observations to a probability distribution over its action space. The policy network is implemented as a three-layer fully connected neural network: 
\begin{equation}
\begin{aligned}
h_1 &= \text{ReLU}(W_1 o^i + b_1), \\
h_2 &= \text{ReLU}(W_2 h_1 + b_2), \\
\pi_{\theta_i}(a \mid o^i) 
&= \text{Softmax}\big(\text{Mask}(W_3 h_2 + b_3)\big),
\end{aligned}
\end{equation}
where $h_1$, $h_2$ denote the hidden-layer representations, and $W$, $b$ are the learnable parameters of the neural network. $\text{ReLU}(x)$ is the activation function, and the Softmax function $\text{Softmax}(x)$ maps the output logits to a normalized probability distribution \cite{nair2010rectified}. The hidden layer dimension is set to 128 to balance representation capacity and computational efficiency. Since different agents may have heterogeneous action spaces, the output layer dimensions vary accordingly, and each agent maintains its own set of policy parameters.

Policy optimization follows the PPO framework. Given sampled trajectories, the clipped surrogate objective is used:
\[
L^{\text{CLIP}}_i = \hat{\mathbb{E}}_k \left[
\min \left(
\rho^i_k(\theta)\hat{A}^i_k,\;
\text{clip}(\rho^i_k(\theta), 1-\epsilon, 1+\epsilon)\hat{A}^i_k
\right)
\right],
\]
where $\rho^i_k(\theta)$ is the probability ratio, $\hat{A}^i_k$ is the estimated advantage and $\epsilon$ is the clipping parameter \cite{yu2022surprising}.

The global value network $V(S_k)$ serves as the critic and it shares the same three-layer MLP backbone with a hidden size of 128, while the output layer is linear to avoid unnecessary range constraints on value prediction. The value network is trained by minimizing the mean squared error between predicted values and target returns:
\[
L^{\text{VALUE}} = \hat{\mathbb{E}}_k \left[
\big(V_{\phi}(S_k) - \hat{R}_k\big)^2
\right],
\]
where $\hat{R}_k$ is computed using generalized advantage estimation \cite{yu2022surprising}.

\begin{figure}[!htbp]
  \centering
  \includegraphics[scale=1]{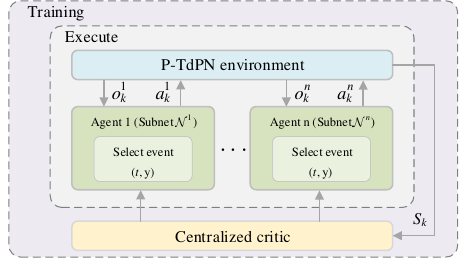}
  \caption{The interaction between the policy and value networks.}
  \label{MAPPO_FIG}
\end{figure}

Fig.~\ref{MAPPO_FIG} illustrates the interaction between the policy and value networks under the centralized training and decentralized execution  framework. During training, the critic evaluates the global state to compute value estimates and advantages, which are used to update the policy of each agent. During execution, only decentralized policies are used for decision-making.

\begin{figure*}[!t]
  \centering
  \begin{subfigure}[t]{0.32\textwidth}
    \centering
    \includegraphics[width=\textwidth]{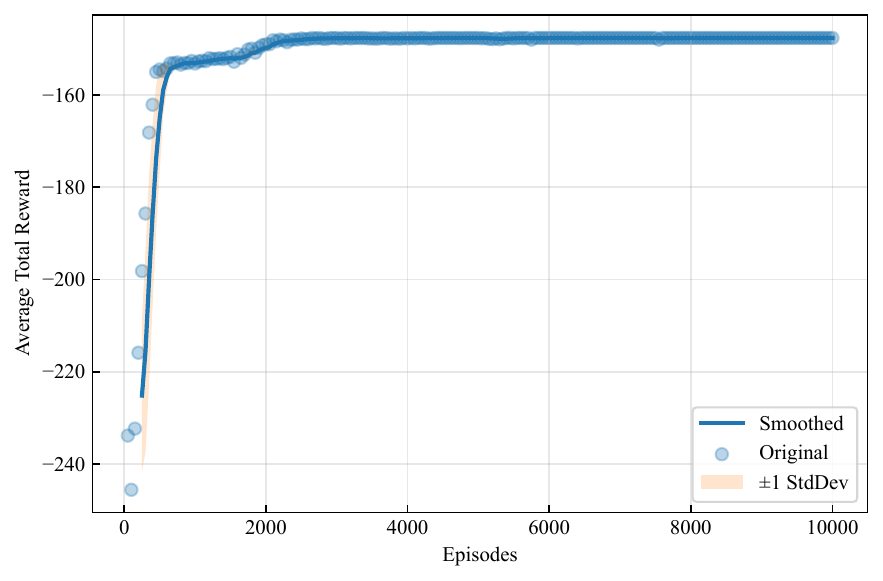}
    \caption{Average total reward}
    \label{subfig1}
  \end{subfigure}
  \hfill
  \begin{subfigure}[t]{0.32\textwidth}
    \centering
    \includegraphics[width=\textwidth]{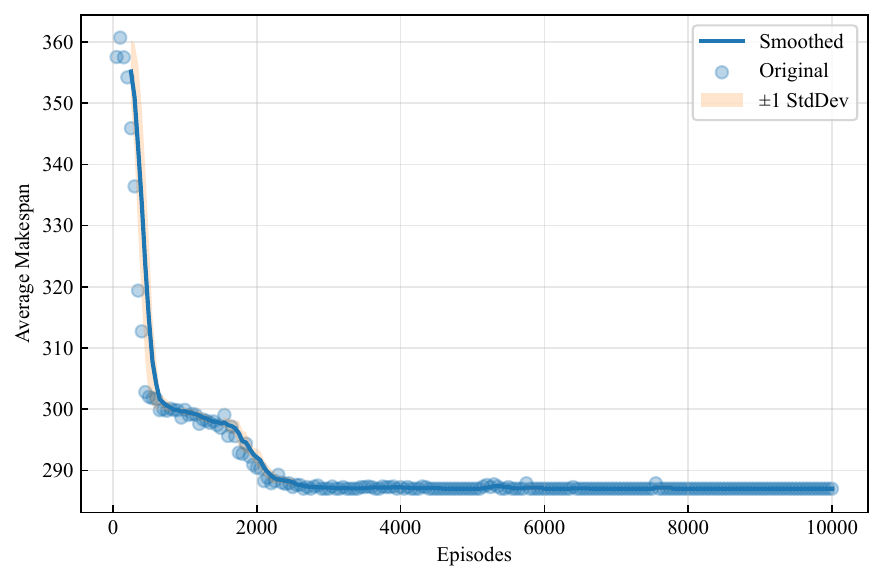}
    \caption{Average Makespan}
    \label{subfig2}
  \end{subfigure}
  \hfill
  \begin{subfigure}[t]{0.32\textwidth}
    \centering
    \includegraphics[width=\textwidth]{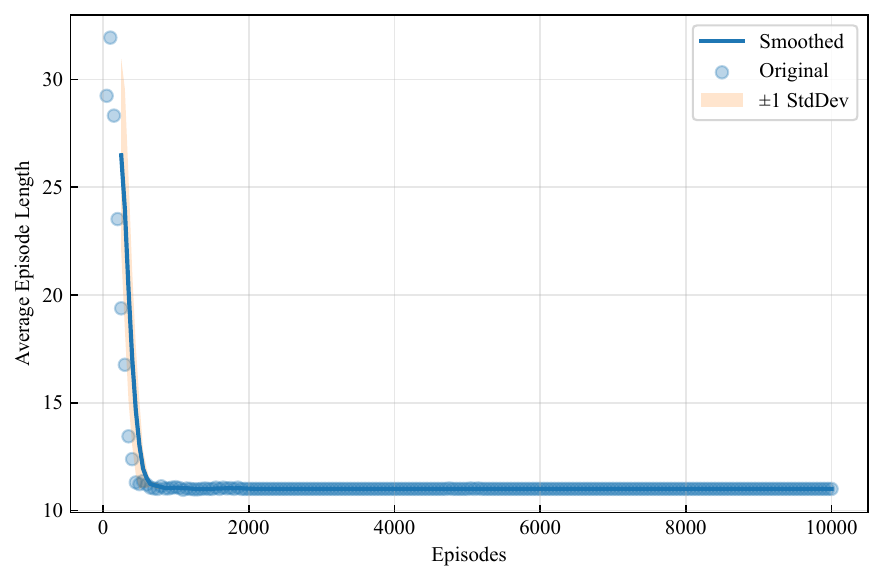}
    \caption{Average episode length}
    \label{subfig3}
  \end{subfigure}
  \caption{Convergence curve during training.}
  \label{Traning_FIG}
\end{figure*}

The overall training procedure is summarized in Algorithm~\ref{MAPPO}. {The P-TdPN serves as the scheduling environment, where basis markings are part of the observation of the agents and events $(t,\mathbf{y})$ correspond to actions. During training, each agent selects a feasible event based on its local observation, while infeasible actions are excluded through action masking. The selected event then drives the environment to evolve to a new basis marking.} At each episode, agents interact with the environment by sampling masked actions based on local observations. The collected trajectories are stored in a buffer and used to compute advantages. The value network is first updated, followed by policy updates for each agent using the proximal policy optimization (PPO) objective. Gradient clipping is applied to stabilize training. The integration of PPO with action masking ensures that all selected actions satisfy PN constraints while maintaining stable and efficient policy optimization.

\begin{algorithm}[!htbp]
\caption{MAPPO based scheduling training with Petri net action masking}
\label{MAPPO}
\KwIn{P-TdPN $\mathcal{N}$, basis partition $\pi=(T_E,T_I)$, maximum episodes $E_{\max}$, number of agents $J$}
\KwOut{Trained policy parameters $\{\theta_i\}_{i=1}^{J}$}
Calculate $Y_c(t)$ for each transition $t \in T_E$ to determine the action space\;
Initialize actor parameters $\{\theta_i\}_{i=1}^{J}$, critic parameters $\phi$\;
\For{$e=1$ \KwTo $E_{\max}$}{
    Reset the environment to inital basis marking $M_0$ and obtain initial observations $S_k$\;
    Initialize replay buffers $\{B_i\}_{i=1}^{J}$\;
    \While{not terminated}{
        \For{$i=1$ \KwTo $J$}{
            Compute mask $v^i$ by screening $Y_c(t)$ against the current basis marking $M_k$\;
            Sample action $a_k^i \sim \pi_{\theta_i}(a_k^i \mid o_k^i,v^i)$\;
        }
        Execute the joint action $(a_k^1,\ldots,a_k^{J})$, which maps to specific BRG events $(t, \mathbf{y})$\;
        Update the environment to the next basis marking $M_{k+1} = M_k + C_I \cdot \mathbf{y} + C(\cdot,t)$\;
        Observe reward $r^i_k$, next observations $S_{k+1}$ from $M_{k+1}$, and termination flags\;
        Store the trajectory into each buffer $B_i$\;
    }
    Compute $V_{\phi}(S_k)$ via the centralized critic\;
    Compute $\hat{A}_k^i$ and $\hat{R}_k$ using generalized advantage estimation\;
    Update the critic $\phi$ by minimizing $L^{\mathrm{VALUE}}$\;
    \For{$i=1$ \KwTo $J$}{
        Compute the probability ratio $\rho^i_k(\theta)$\;
        Update $\theta_i$ by maximizing $L_{i}^{\mathrm{CLIP}}$\;
    }
}
\Return $\{\theta_i\}_{i=1}^{J}$\;
\end{algorithm}

\section{Experimental Results}\label{5}
The proposed algorithm is implemented in Python and executed on a personal computer equipped with a 2.1\,GHz processor and 16\,GB of RAM. In the subsequent subsections, several benchmark instances are evaluated, and the results are compared with those obtained from existing methods. The source code corresponding to all experimental cases is publicly accessible online\footnote{https://github.com/FMS-Scheduling-0831/FMS-Dynamic-Scheduling}.

\subsection{Illustrative Example}
To validate the effectiveness of the proposed algorithm, the P-TdPN model of the FMS shown in Fig.~\ref{PTPN_Indus} is used as a case study. The initial marking is given as $M_0 = [5,0,0,0,0,0,5,0,0,0,0,1,1,1,1]^{T}$, and the basis partition is defined as $\pi = (T_E, T_I)$, where $T_I = \{t_{1}, t_{4}, t_{5}, t_{7}, t_{9}\}$ and $T_E = T \setminus T_I$. Simulation experiments are conducted using the MAPPO-based FMS scheduling algorithm. 

It should be noted that the parameters employed in the algorithm were selected based on preliminary testing using the following instances. The algorithm parameters are configured as follows: the Adam optimizer is adopted for both the actor and critic networks, with learning rates of $3 \times 10^{-4}$ and $1 \times 10^{-3}$, respectively. The clipping parameter is set to $\epsilon = 0.2$, and the generalized advantage estimation parameter is set to $0.95$. A step penalty of $-10$ is imposed. The total number of training episodes is set to 10,000, and the maximum number of interaction steps per episode is determined according to the problem scale to prevent unbounded interactions. During training, three key metrics—the average total reward, average episode length, and average makespan—are recorded every 50 episodes. A moving average method is applied to smooth the raw data and better illustrate the training trends.

The average cumulative reward (Fig.~\ref{subfig1}) exhibits a three-stage convergence pattern, increasing from $-250$ to $-160$ within the first 200 episodes, followed by a gradual improvement to $-149$ over episodes 200--2000, and finally stabilizing at $-147 \pm 2$ after 2000 episodes, with fluctuations below $1.5\%$. The average episode length (Fig.~\ref{subfig2}) decreases from approximately 33 to 11 steps (a 66\% reduction), with the most significant decline occurring within the first 200 episodes and convergence achieved after 800 episodes with near-zero variance, indicating efficient multi-agent coordination and reduced exploration complexity. The average makespan (Fig.~\ref{subfig3}) is reduced from 361 to 287 (approximately 20\%), converging to 287 after 1200 episodes, which suggests that the learned policy is both near-optimal and stable.

\iffalse
\begin{figure*}[!t]
  \centering
  \begin{subfigure}[t]{0.32\textwidth}
    \centering
    \includegraphics[width=\textwidth]{myfigure/Average_Episode_Length_comparison.pdf}
    \caption{The impact on episode length}
    \label{subfig1}
  \end{subfigure}
  %\hfill
  \hspace{-5pt}
  \begin{subfigure}[t]{0.32\textwidth}
    \centering
    \includegraphics[width=\textwidth]{myfigure/Average_Makespan_comparison.pdf}
    \caption{The impact on Makespan}
    \label{subfig2}
  \end{subfigure}
  \hspace{-5pt}
  \begin{subfigure}[t]{0.32\textwidth}
    \centering
    \includegraphics[width=\textwidth]{myfigure/Success_Rate_comparison.pdf}
    \caption{The impact on success rate}
    \label{subfig3}
  \end{subfigure}
  \caption{Ablation study on the impact of reward components.}
  \label{Traning_FIG}
\end{figure*}
\fi

To assess the practical scheduling performance of the converged policy, we deploy the trained actors to perform inference on the test instance. Table~\ref{decision} summarizes the step-by-step scheduling decisions generated in one complete rollout. The policy completes the entire job schedule within 11 steps and achieves a final makespan of 287. Notably, in steps 1-2 and 4-10, both agents fire transitions on their respective production lines simultaneously; only steps 3 and 11 are executed by a single agent. This behavior indicates that the decentralized policies learned under the centralized training and decentralized execution paradigm capture a highly coordinated parallel scheduling strategy, keeping the two production lines synchronized for most of the horizon and thereby improving overall processing efficiency.
\begin{table}[!htbp]
\centering
\small
\renewcommand{\arraystretch}{1.2}
\caption{Scheduling decision of MAPPO algorithm.}
\label{decision}
\begin{tabular}{ccccc}
\toprule
\multirow{2}{*}{\raisebox{-0.3em}{Step}}
& \multicolumn{2}{c}{Agent 1} 
& \multicolumn{2}{c}{Agent 2} \\
\cmidrule(lr){2-3} \cmidrule(lr){4-5}
& Explicit & Implicit & Explicit & Implicit \\
\midrule
01 & $t_{2}$ & $t_{1}$ & $t_{8}$ & $t_{7}$ \\
02 & $t_{6}$ & $t_{4}$ & $t_{8}$ & $t_{7}, t_{9}$ \\
03 & --      & --      & $t_{10}$& -- \\
04 & $t_{2}$ & $t_{1}$ & $t_{8}$ & $t_{7}, t_{9}$ \\
05 & $t_{6}$ & $t_{4}$ & $t_{10}$& -- \\
06 & $t_{2}$ & $t_{1}$ & $t_{10}$& $t_{9}$ \\
07 & $t_{6}$ & $t_{4}$ & $t_{8}$ & $t_{7}$ \\
08 & $t_{2}$ & $t_{1}$ & $t_{8}$ & $t_{7}, t_{9}$ \\
09 & $t_{6}$ & $t_{4}$ & $t_{10}$& -- \\
10 & $t_{2}$ & $t_{1}$ & $t_{10}$& $t_{9}$ \\
11 & $t_{6}$ & $t_{4}$ & --      & -- \\
\bottomrule
\end{tabular}
\end{table}

Table~\ref{table_diff} reports the scheduling results obtained under randomly generated different basis partitions. Note that $T_I = \emptyset$ indicates 
that no implicit transitions are selected, i.e., the proposed approach is tested based on the RG (without the state space compression of BRG). It can be observed that most partitions achieve a makespan of either 287 or 293, indicating that the scheduling performance is relatively insensitive to the specific choice of basis partition. 
Notably, basis partitions with larger implicit transition sets generally require fewer decision steps while achieving comparable makespans. This result indicates that increasing the number of implicit transitions can further compress the BRG representation and improve decision efficiency without significantly affecting scheduling quality. As observed, the  RG approach yields the worst makespan of 323. Without state abstraction, the elongated Markov decision horizon severely complicates temporal credit assignment, causing the RL agent to prematurely converge to suboptimal policies.

%In contrast, the number of decision steps varies across partitions, ranging from 11 to 21. This difference arises because larger implicit transition sets allow longer implicit firing sequences to be aggregated into BRG arcs, thereby reducing the number of basis markings visited during scheduling. 

\begin{table}[!htbp]
\centering
\renewcommand{\arraystretch}{1.2}
\caption{Scheduling results for different basis partitions.}
\scalebox{1.2}{
\begin{tabular}{ccc}
\hline
$T_I$  & Steps    & Makespan \\ \hline
$\emptyset$   & 20  & 323 \\ 
%$\{t_{3}, t_{4}, t_{8}\}$                  & 21  & 287 \\
$\{t_{2}, t_{3}, t_{7}\}$                  & 20  & 287 \\
$\{t_{1}, t_{4}, t_{8}\}$                  & 15  & 293 \\
$\{t_{2}, t_{5}, t_{7}, t_{9}\}$           & 16  & 293 \\
$\{t_{1}, t_{6}, t_{7}, t_{9}\}$           & 16  & 287 \\
$\{t_{3}, t_{4}, t_{8}, t_{10}\}$          & 18  & 287 \\
$\{t_{3}, t_{4}, t_{7}, t_{10}\}$          & 16  & 293 \\
$\{t_{1}, t_{4}, t_{5}, t_{7}, t_{9}\}$    & 11  & 287 \\
%$\{t_{2}, t_{}, t_{5}, t_{7}, t_{10}\}$   & 12  & 293 \\ 
\hline
\end{tabular}}
\label{table_diff}
\end{table}

\subsection{Ablation Experiments}
\begin{figure}[!htbp]
  \centering
  \includegraphics[scale=0.4]{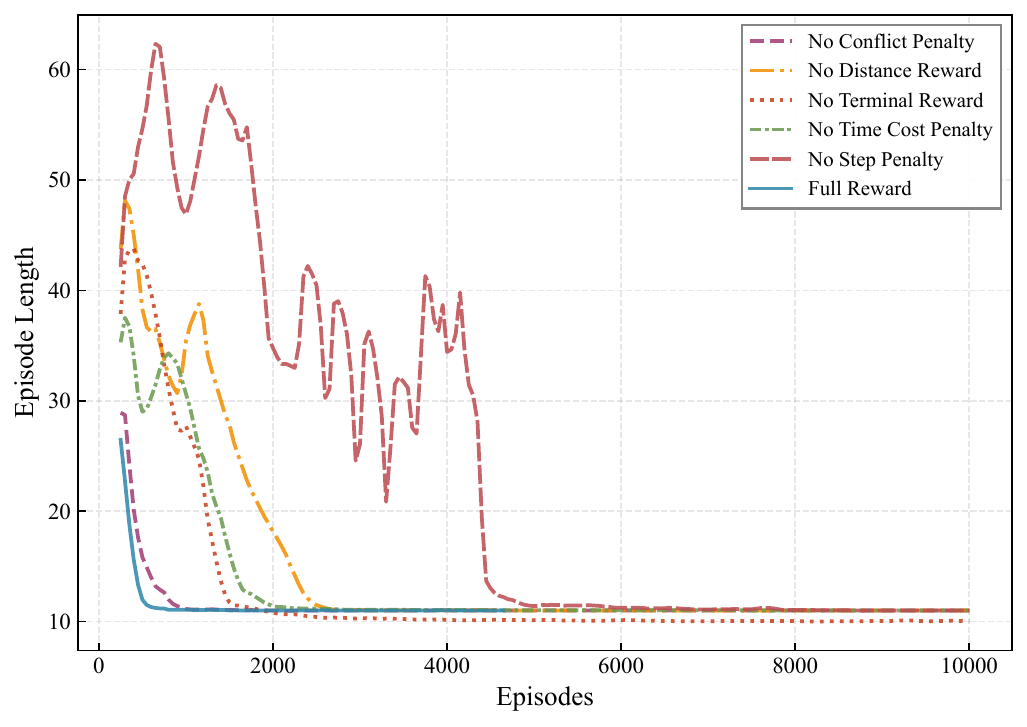}
  \caption{Study of reward components on episode length.}
  \label{Length}
\end{figure}

\begin{figure}[!htbp]
  \centering
  \includegraphics[scale=0.4]{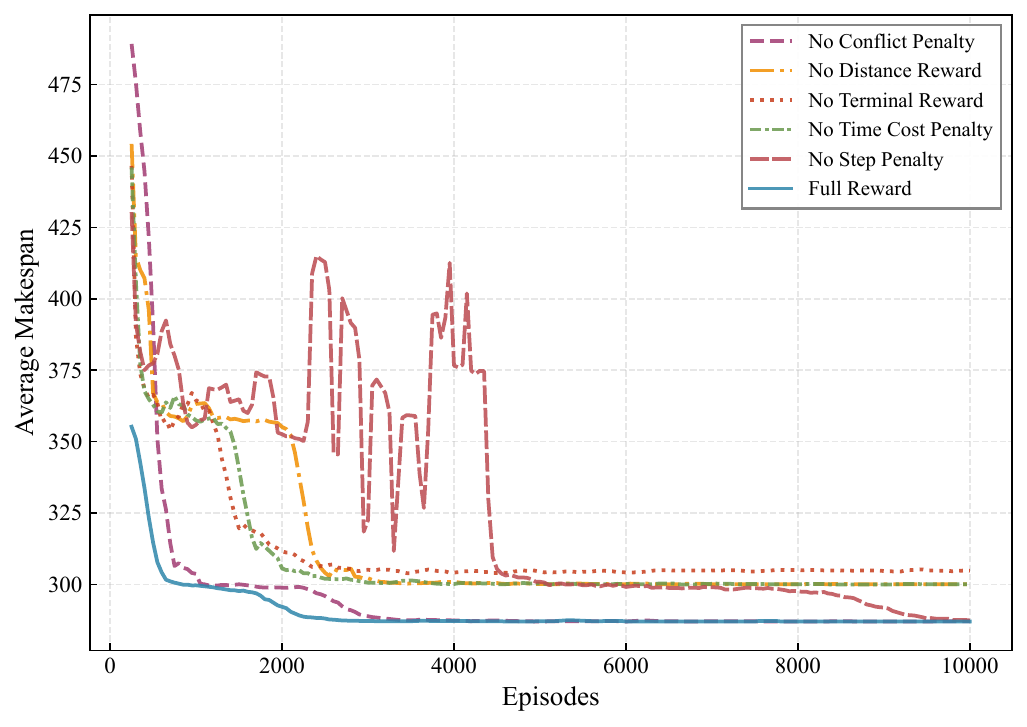}
  \caption{Study of reward components on Makespan.}
  \label{Makespan}
\end{figure}

\begin{figure}[!htbp]
  \centering
  \includegraphics[scale=0.4]{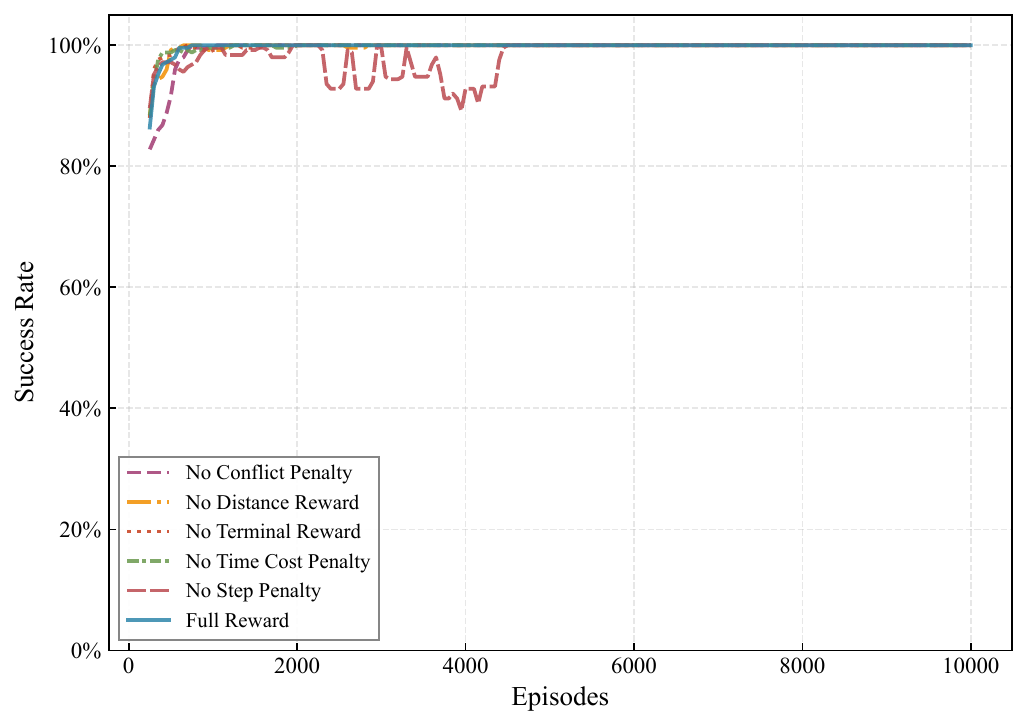}
  \caption{Study of reward components on success rate.}
  \label{Success}
\end{figure}

To validate the contribution of each reward component, we conduct ablation experiments by removing one component at a time. As illustrated in Figs.~\ref{Length}, \ref{Makespan} and \ref{Success}, all configurations ultimately achieved a 100\% success rate (defined as completing the scheduling task within a limited number of steps), indicating that task completion is robust to variations in reward design. However, substantial differences are observed in convergence speed, training stability, and scheduling quality.

Step penalties constitute the most critical component for ensuring training stability. Their removal results in pronounced oscillatory behavior during the early stages of training, with episode lengths fluctuating between 12 and 60 steps across multiple cycles, as shown in Fig.~\ref{Length}. This suggests that per-step penalties are essential to prevent policies from degenerating into inefficient random exploration and to encourage agents to identify shorter paths for task completion.

The terminal reward has the most significant impact on scheduling quality, as its removal consistently leads to an increased makespan, as shown in Fig.~\ref{Makespan}. This confirms that explicitly penalizing the cumulative scheduling time at episode completion provides a crucial signal for optimizing makespan. While the distance-based reward and time cost penalty primarily enhance convergence efficiency and scheduling quality. Conflict penalties improve the success rate during the early stages of training and facilitate coordination among agents, as shown in Fig.~\ref{Success}.

The full reward configuration achieves the fastest convergence, the lowest makespan, and the most stable training trajectory, validating that each component contributes complementarily to the overall learning objective.

\subsection{Results on Benchmark Datasets}
{Dynamic events alter the marking evolution of the PN, resulting in changes to the system state. Since the current marking is encoded in the state representation and provided as input to the trained agents, the agents can adapt their decisions according to the system states. Consequently, the proposed method is capable of responding to dynamic events.}

To comprehensively evaluate the proposed approach for online dynamic scheduling under FMSs of different scales, we adopt the benchmark instances FMS01--FMS20 from \cite{Han2015} and In01--In16 from \cite{KeYiXing2012}. Based on these static instances, we construct two dynamic test suites, DFMS01--DFMS20 and DIn01--DIn16, by introducing mixed dynamic events. Specifically, new jobs arrive randomly during processing, where the number of arrivals is sampled as a random integer in $[1, 5]$; machine breakdowns occur stochastically, and both the time-to-failure and the repair time follow exponential distributions; to emulate time-varying customer demands, jobs may be cancelled online, where the cancellation time follows an exponential distribution and exactly one job is cancelled per event.

We compare our method with four widely used online dispatching rules, namely First-In-First-Out (FIFO), Shortest Processing Time (SPT), Longest Processing Time (LPT), and a Greedy heuristic (GREEDY) \cite{Lee2009}. In addition, to demonstrate the advantages of using BRG based MAPPO, we also tested the RG based MAPPO method and employed random basis partition in the BRG according to \cite{Ma2017}. Each algorithm was run 20 times, and Table~\ref{DFMS} reports the average makespan results on DFMS01--DFMS20.

\begin{table}[!htbp]
\centering
\small
\renewcommand{\arraystretch}{1.2}
\caption{Scheduling results of examples DFMS01--20.}
\label{DFMS}
\scalebox{0.8}{
\begin{tabular}{ccccccc}
\toprule
\multirow{3}{*}{Instance} & \multicolumn{6}{c}{Makespan} \\
\cline{2-7}
 & {\raisebox{-0.2em}{MAPPO}} & {\raisebox{-0.2em}{MAPPO}} & {\raisebox{-0.2em}{}} & {\raisebox{-0.2em} {}}
 & {\raisebox{-0.2em}{}}   & {\raisebox{-0.2em}{}} \\
 & {\raisebox{-0.2em}{(BRG)}} & {\raisebox{-0.2em}{(RG)}} & {\raisebox{-0.2em}{SPT}} & {\raisebox{-0.2em}{FIFO}} 
 & {\raisebox{-0.2em}{GREEDY}}   & {\raisebox{-0.2em}{LPT}} \\
\midrule
DFMS01 & \textbf{386.2}  & 407.7  & 432.8  & 445.2  & 476.3  & 481.5 \\
DFMS02 & \textbf{664.9}  & 682.2  & 712.7  & 735.6  & 743.1  & 774.2 \\
DFMS03 & \textbf{1184.2} & 1211.6 & 1233.9 & 1254.8 & 1247.6 & 1262.4 \\
DFMS04 & \textbf{1712.1} & 1731.3 & 1754.1 & 1768.8 & 1766.2 & 1781.6 \\
DFMS05 & \textbf{2754.6} & 2781.4 & 2781.4 & 2822.9 & 2817.4 & 2843.1 \\
DFMS06 & \textbf{211.3}  & 228.6  & 249.1  & 254.7  & 263.3  & 271.2 \\
DFMS07 & \textbf{318.7}  & 334.9  & 352.3  & 373.0  & 348.2  & 382.5 \\
DFMS08 & \textbf{607.2}  & 621.7  & 638.1  & 659.4  & 682.7  & 691.8 \\
DFMS09 & \textbf{864.5}  & 877.1  & 893.2  & 911.3  & 882.5  & 925.3 \\
DFMS10 & \textbf{1377.4} & 1392.1 & 1413.6 & 1438.5 & 1425.1 & 1456.7 \\
DFMS11 & \textbf{154.3}  & 176.2  & 191.5  & 210.2  & 204.6  & 223.8 \\
DFMS12 & \textbf{221.6}  & 235.5  & 254.3  & 269.4  & 263.1  & 277.1 \\
DFMS13 & \textbf{390.0}  & 413.2  & 439.7  & 441.4  & 427.4  & 447.6 \\
DFMS14 & \textbf{552.1}  & 568.9  & 572.8  & 591.4  & 584.2  & 605.3 \\
DFMS15 & \textbf{941.2}  & 962.6  & 974.8  & 981.3 & 995.7 & 1009.0 \\
DFMS16 & \textbf{133.5}  & 152.3  & 167.3  & 174.7  & 185.2  & 197.3 \\
DFMS17 & \textbf{193.4}  & 212.4  & 225.7  & 256.9  & 231.6  & 271.1 \\
DFMS18 & \textbf{307.0}  & 315.6  & 323.1  & 352.7  & 331.3  & 345.8 \\
DFMS19 & \textbf{452.1}  & 471.9  & 485.2  & 498.6  & 492.2  & 507.4 \\
DFMS20 & \textbf{691.3}  & 716.8  & 727.5  & 743.1  & 735.0  & 761.2 \\
\bottomrule
\end{tabular}}
\end{table}
Our method achieves the smallest makespan in all 20 instances, demonstrating consistent and statistically stable superiority over heuristic baselines, improving upon RG based MAPPO, SPT, FIFO, GREEDY, and LPT by 4.51\%, 8.14\%, 11.36\%, 10.52\%, and 14.09\%, respectively. It is worth mentioning that, compared with RG, BRG reduces the action search space, alleviates resource conflicts, and decreases the occurrence of deadlock states. Consequently, agents can explore more efficiently and learn higher quality scheduling policies, resulting in improved scheduling performance. Moreover, once training is completed, policy inference is highly efficient: the online decision time is below 0.5\,s.

\begin{table}[!htbp]
\centering
\small
\renewcommand{\arraystretch}{1.2}
\caption{Scheduling results of examples DIn01--16.}
\label{DIn}
\scalebox{0.8}{
\begin{tabular}{ccccccc}
\toprule
\multirow{3}{*}{Instance} & \multicolumn{6}{c}{Makespan} \\
\cline{2-7}
 & {\raisebox{-0.2em}{MAPPO}} & {\raisebox{-0.2em}{MAPPO}} & {\raisebox{-0.2em}{}} & {\raisebox{-0.2em} {}}
 & {\raisebox{-0.2em}{}}   & {\raisebox{-0.2em}{}} \\
 & {\raisebox{-0.2em}{(BRG)}} & {\raisebox{-0.2em}{(RG)}} & {\raisebox{-0.2em}{SPT}} & {\raisebox{-0.2em}{FIFO}} 
 & {\raisebox{-0.2em}{GREEDY}}   & {\raisebox{-0.2em}{LPT}} \\
\midrule
DIn01 & \textbf{331.6} & 346.4 & 361.1 & 388.9 & 382.1 & 393.3 \\
DIn02 & \textbf{405.2} & 420.7 & 443.3 & 464.1 & 457.6 & 481.5 \\
DIn03 & \textbf{513.7} & 531.2 & 545.3 & 561.7 & 573.5 & 587.4 \\
DIn04 & \textbf{584.4} & 599.6 & 611.1 & 637.4 & 622.3 & 658.7 \\
DIn05 & \textbf{322.1} & 336.0 & 351.9 & 372.1 & 365.3 & 389.6 \\
DIn06 & \textbf{387.6} & 401.2 & 425.3 & 455.9 & 452.7 & 463.2 \\
DIn07 & \textbf{501.3} & 522.1 & 545.6 & 558.7 & 538.2 & 561.4 \\
DIn08 & \textbf{572.9} & 586.3 & 593.4 & 617.5 & 624.6 & 639.2 \\
DIn09 & \textbf{299.6} & 315.4 & 337.1 & 351.9 & 362.3 & 375.7 \\
DIn10 & \textbf{368.7} & 379.8 & 392.3 & 416.1 & 409.8 & 425.3 \\
DIn11 & \textbf{452.5} & 463.4 & 477.1 & 496.6 & 481.4 & 510.9 \\
DIn12 & \textbf{554.1} & 568.5 & 581.3 & 601.9 & 593.2 & 616.7 \\
DIn13 & \textbf{251.8} & 264.2 & 282.4 & 295.8 & 307.5 & 318.1 \\
DIn14 & \textbf{300.3} & 315.7 & 331.1 & 347.2 & 339.1 & 356.7 \\
DIn15 & \textbf{332.2} & 347.1 & 369.3 & 378.9 & 371.2 & 385.5 \\
DIn16 & \textbf{503.4} & 516.4 & 530.7 & 552.1 & 547.8 & 561.3 \\
\bottomrule 
\end{tabular}}
\end{table}

Table~\ref{DIn} summarizes the results on DIn01--DIn16, which have more production lines and resource types than DFMS01-DFMS20. The experimental results demonstrate that, as the scale of the problem increases and the complexity grows, the proposed method again yields the best makespan in all 16 instances, with an average improvement ratio of 9.43\%.

\section{Conclusion and Future Work}\label{conclusion}
This paper proposes a dynamic scheduling method for FMS based on multi-agent deep reinforcement learning and Petri nets. Based on a compact state space representation structure, the action space and state space of the MDP are designed, and an MAPPO algorithm is proposed for dynamic FMS scheduling.  Experiments on practical benchmark
systems show that our method significantly reduces the makespan compared to common rule-based approaches, while maintaining real-time responsiveness for dynamic events.  Future research will focus on extending the approach to handle more complex production scenarios, including multi-batch processing, as well as integrating resource and energy constraints into the scheduling framework.

\bibliographystyle{IEEEtran}
\bibliography{IEEEtest}

\end{document}